\documentclass[twocolumn,showpacs,preprintnumbers,amsmath,amssymb,prb]{revtex4}

\usepackage{color}
\usepackage{graphicx}
\usepackage{dcolumn} 
\usepackage{mathrsfs}
\usepackage{bm}
\usepackage{bbm}
\usepackage{overpic}
\usepackage[latin1]{inputenc}
\usepackage[T1]{fontenc}
\usepackage{amsmath}
\usepackage{amsfonts}
\usepackage{amstext}     
\usepackage{amssymb}
\usepackage{epsfig}

\begin{document}

\title{Floquet topological quantum phase transitions in the transverse Wen-plaquette model}
\author{V. M. Bastidas$^1$}
\email{victor@physik.tu-berlin.de}
\author{C. Emary$^1$}
\author{G. Schaller$^1$}
\author{A. G\'omez-Le\'on$^2$}
\author{G. Platero$^2$}
\author{T. Brandes$^1$}
\affiliation{%
$^1$Institut f\"ur Theoretische Physik, Technische Universit\"at Berlin, Hardenbergstr. 36, 10623 Berlin, Germany}%
\affiliation{%
$^2$Instituto de Ciencia de Materiales de Madrid (ICMM-CSIC), Cantoblanco, 28049 Madrid, Spain}
%
\begin{abstract}
Our aim in this work is to study the nonequilibrium behavior of the topological quantum phase transition in the transverse Wen-plaquette model.
We show that under the effect of a nonadiabatic driving the system exhibits a new topological phase and a rich phase diagram. We define generalized topological order parameters by considering cycle-averaged expectation values of string operators in a Floquet state.
\end{abstract}


\maketitle
\section{Introduction \label{SectionO}} 
Nature admits transitions between different states of matter driven by thermal fluctuations,
which are referred to as classical phase transitions \cite{Landau1,Landau2}.
At zero temperature, as is expected, the thermal fluctuations do not play
any role, however, the quantum fluctuations can entirely drive a change of phase in the system, which  corresponds to a quantum phase transition (QPT). In contrast to classical phase transitions, a QPT is characterized by a dramatic change
of the ground state properties \cite{Sachdev}.

Symmetry breaking is a paradigm of condensed matter physics in which the states of matter are characterized as symmetry broken phases \cite{Landau1, Landau2}. Rather recently, novel states of matter have been found which cannot be classified inside the paradigm of Landau symmetry breaking \cite{Kitaev1,Kitaev2,WenPRL}. These states of matter are characterized by an intrinsic robustness against the effects of an environment as a consequence of the topological properties of the ground state. 
Examples of such topological states include the quantum Hall effect \cite{VonKlitzing,Aoki,denNijs}, 
topological insulators and topological superconductors \cite{Kitaev3, ReviewHasan, ReviewZhang}. This manifestation of criticality in nature is characterized by transitions between different topological quantum numbers, and corresponds to a topological quantum phase transition (TQPT). 
In the integer quantum Hall effect, the different topological phases correspond to the Hall Plateaus observed experimentally \cite{VonKlitzing}. A transition between states with different Hall conductance can occur without symmetry breaking. Importantly, further investigations in  spin systems with topological order will have a huge potential for applications in quantum information as topologically-protected qubits \cite{Kitaev2}. 

Recent theoretical studies predicted the existence of topological insulators in solid state systems \cite{ReviewHasan, ReviewZhang,Bernevig}, with a subsequent experimental observation in semiconductor heterostructures \cite{Koenig,ChenThreeDTop}. Further investigations describe the emergence of Majorana modes at the edges of quantum wires \cite{Lutchyn,Oreg}. However, the experimental observation of the Majorana resonance is still under debate \cite{Mourik,Alicea}.

Despite of the striking results on TQPT in solid state systems, there is an increasing interest
on the realization of topological states in other experimental setups. 
In particular, a realization of quantum magnets in a system of cold atoms placed in an optical lattice has been suggested recently \cite{Duan,ZollerK,ZollerTC}, to allow the implementation of the spin-$1/2$ Kitaev model on a hexagonal lattice. Furthermore, topological protection can be realized by means of an array of superconducting nanocircuits \cite{Gershenson}, which is characterized by an $Z_2$ topological order parameter \cite{WenPRL}. 

Previous investigations in QPTs inside the Landau symmetry-breaking paradigm, show that the external control induces effective interactions that lead to the existence of
new quantum phases \cite{Bastidas3,Bastidas4,BastidasLMG}. A characteristic of the driven manybody systems is that the periodicity in time opens the possibility to create gapless excitations in a controlled way by tuning the parameters around quantum resonances \cite{Tanaka,HolthausQPT, Vedral, Creffield}. Furthermore, the influence of the external control on the topology of the system via geometric phases has also been explored \cite{PlateroBerry,TomkaPolkov}.

The current experimental feasibilities have motivated the study of TPQTs under an external driving. Rather recently, the concept of topological charge of the Floquet Majorana fermions defined in terms of the Floquet operator has been introduced \cite{ZollerTopo}. 
Furthermore, under the effect of intense circularly polarized light, a gap can be open in the Dirac cone, leading to a photoinduced dc Hall current in graphene \cite{AokiTanaka}.
Similarly, by using one-photon resonances, a trivial insulator can be driven into the topological phase,  originating a Floquet topological insulator \cite{AokiTanaka,Lindner}. 

In contrast to the works previously mentioned, in this paper we show that the driving not only allows us to renormalize the parameters, such that the system enters into the topological phases, but, surprisingly, it induces a new topological phase that is absent in the undriven system. 
In this work we study the Wen-plaquete model (WPM) in a monochromatic transverse field. The WPM has intriguing relations to other models studied in the literature. For example, it has been shown that the WPM can be exactly mapped to the toric code of Kitaev \cite{Nussinov,BrownVedral}. Furthermore, the toric code in a paralell magnetic field has low energy properties that resemble the two-dimensional transverse Ising model, which allows to study the influence of an external field in the TQPT \cite{Vidal1,VidalPRL}. The effect of a transverse perturbation on the topological protection in the toric code has been explored by using a mapping onto the Xu-Moore model, which subsequently can be
mapped onto the quantum compass model \cite{Vidal2}.

The paper is organized as follows: In Sec. \ref{SectionI} we discuss the general aspects of the transverse Wen-plaquette model (TWPM). In Sec. \ref{SectionII} we describe the physics of the system by means of the rotating wave approximation (RWA) and discuss signatures of criticality based on the description of the nonequilibrium Ising transition. In Sec. \ref{SectionIII} we define generalized ``string''-like topological order parameters by considering cycle-averaged expectation values of string operators in a Floquet state. Finally, a discussion of the results is presented in Sec. \ref{SectionIV}.

\section{Generalities of the transverse Wen-plaquette model \label{SectionI}} 
\begin{figure}
\centering
\includegraphics[width=0.45 \textwidth]{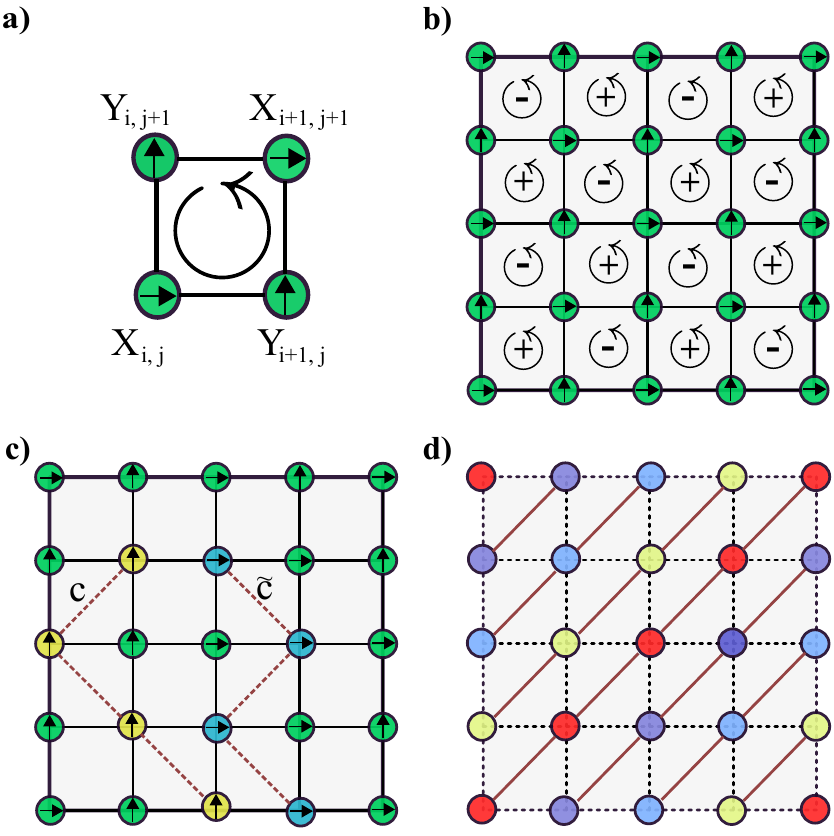}
\caption{
  \label{FigW1}
      The transverse Wen-plaquette model. In (a) we represent the geometry of plaquette operator at the site $(i,j)$. Figure (b) depicts the real space lattice and the corresponding even $(+)$ and odd $(-)$ plaquettes. (c) Shows the $T_1$ string operator along the path $C$, and the $T_2$ string operator along $\widetilde{C}$. Figure (d) shows the dual lattice and represents the $N$ decoupled Ising chains. To take into account the periodic boundary conditions, in the dual lattice the colors represent the different Ising chains. 
       } 
\end{figure}

In this section we describe the driven Wen-plaquette model in a transverse field on a $N\times N$ square lattice
\begin{eqnarray}
      \label{drivenWen}
            \hat{H}(t)&=& -g(t)\sum_{i,j}\hat{X}_{i,j}-J\sum_{i,j} \hat{F}^{z}_{i,j}
      ,
\end{eqnarray}
where
\begin{eqnarray}
      \label{Plaquetteoperator}
            \hat{F}^{z}_{i,j} = \hat{X}_{i,j} \hat{Y}_{i+1,j}
            \hat{X}_{i+1,j+1} \hat{Y}_{i,j+1}
\end{eqnarray}
is the plaquette operator, and $g(t)=g_{0}+g_{1}\cos\Omega t$. In this paper, $\hat{X}_{i,j}$,$\hat{Y}_{i,j}$, and  $\hat{Z}_{i,j}$ denote the Pauli operators acting on the $(i,j)$-th site of the square lattice. Furthermore, we assume periodic boundary conditions for even $N$.
Fig.~\ref{FigW1} $(a)$ depicts the geometry of the plaquette operator. A plaquette is even (odd) if the relation $(-1)^{i + j} = 1$ $[(-1)^{i + j} = -1]$ is satisfied \cite{Wen2}, as depicted in Fig.~\ref{FigW1} $(b)$.

In the topological phase, the undriven TWPM accommodates three types of nonlocal excitations above the ground state of the system: $Z_2$ charges,
$Z_2$ vortices, and fermions. Correspondingly, to describe these quasiparticles, the string operators of types $T_1$, $T_2$, and $T_3$ are defined \cite{Wen2}. The $T_1$ string operator for a $Z_2$ charge is given by 
\begin{eqnarray}
      \label{Z2Charge}
            \widehat{W}_{\mathrm{c}}(C)=\bigotimes_C \hat{Y}_{i,j}
      .
\end{eqnarray}
$\widehat{W}_{\mathrm{c}}$ is the product of spin operators along a path $C$ connecting even plaquettes of neighboring links (see Fig.~\ref{FigW1} (c) ). Correspondingly, the $T_2$ string operator for a $Z_2$ vortex is given by
\begin{eqnarray}
      \label{Z2Vortex}
             \widehat{W}_{\mathrm{v}}(\widetilde{C})=\bigotimes_{\widetilde{C}} \hat{X}_{i,j}
      .
\end{eqnarray}
In this case, $\widehat{W}_{\mathrm{v}}$ is the product of spin operators along a path $\tilde{C}$ connecting odd plaquettes of neighboring links (see Fig.~\ref{FigW1} (c)). Finally, the $T_3$ string operators are defined as bound states of $T_1$ and $T_2$ strings, they are charge-vortex composite objects, and their definition is given in Refs. \cite{Wen2,Wen1}.
\subsection{The duality transformation \label{SubSectionI1}} 
In this section, following the methods of Refs. \cite{BrownVedral,Wen2,Wen1}, we show that the driven TWPM Hamiltonian Eq. \eqref{drivenWen} can be mapped into a set of decoupled
one-dimensional Ising chains via a duality transformation.

\begin{figure}
\centering
\vspace{0.4cm}
  \begin{minipage}[b]{0.491\linewidth}
    \begin{overpic}[width=\linewidth]{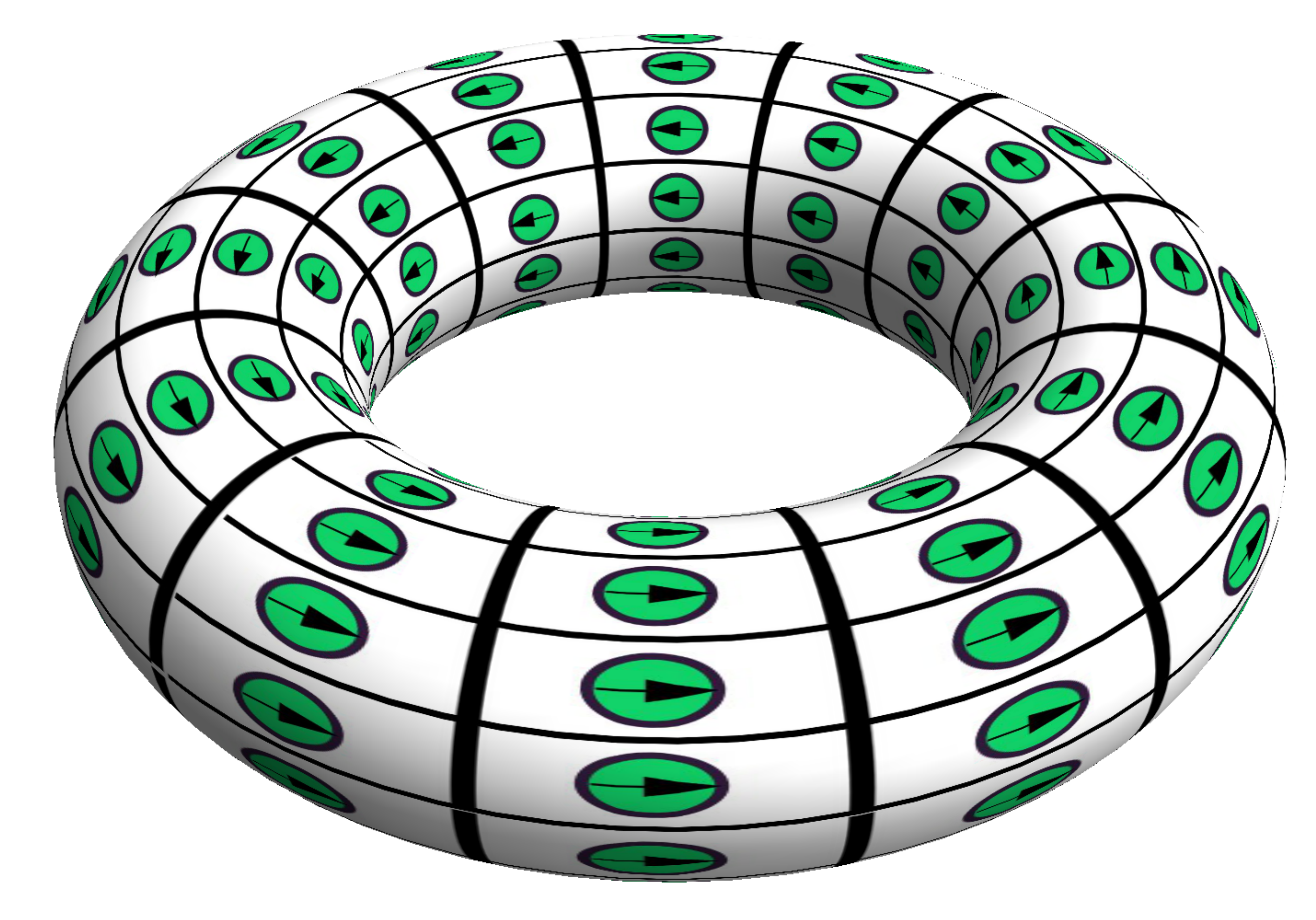}
    \put(-2,60){ \textbf a)}
    \end{overpic}
  \end{minipage}
  \begin{minipage}[b]{0.491\linewidth}
    \begin{overpic}[width=\linewidth]{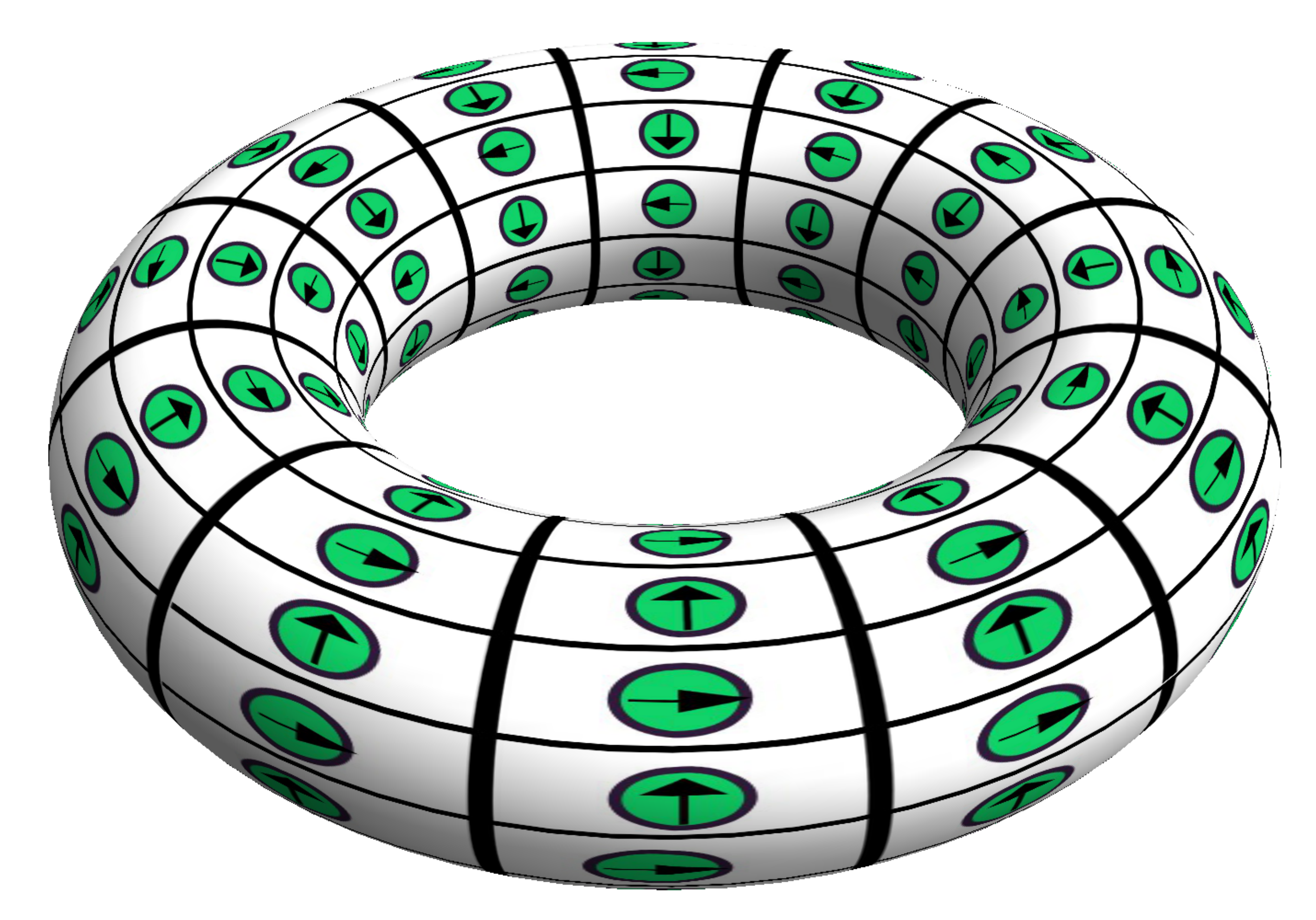}
    \put(-2,60){ \textbf b)}
    \end{overpic}
  \end{minipage}
   \begin{minipage}[b]{0.491\linewidth}
    \begin{overpic}[width=\linewidth]{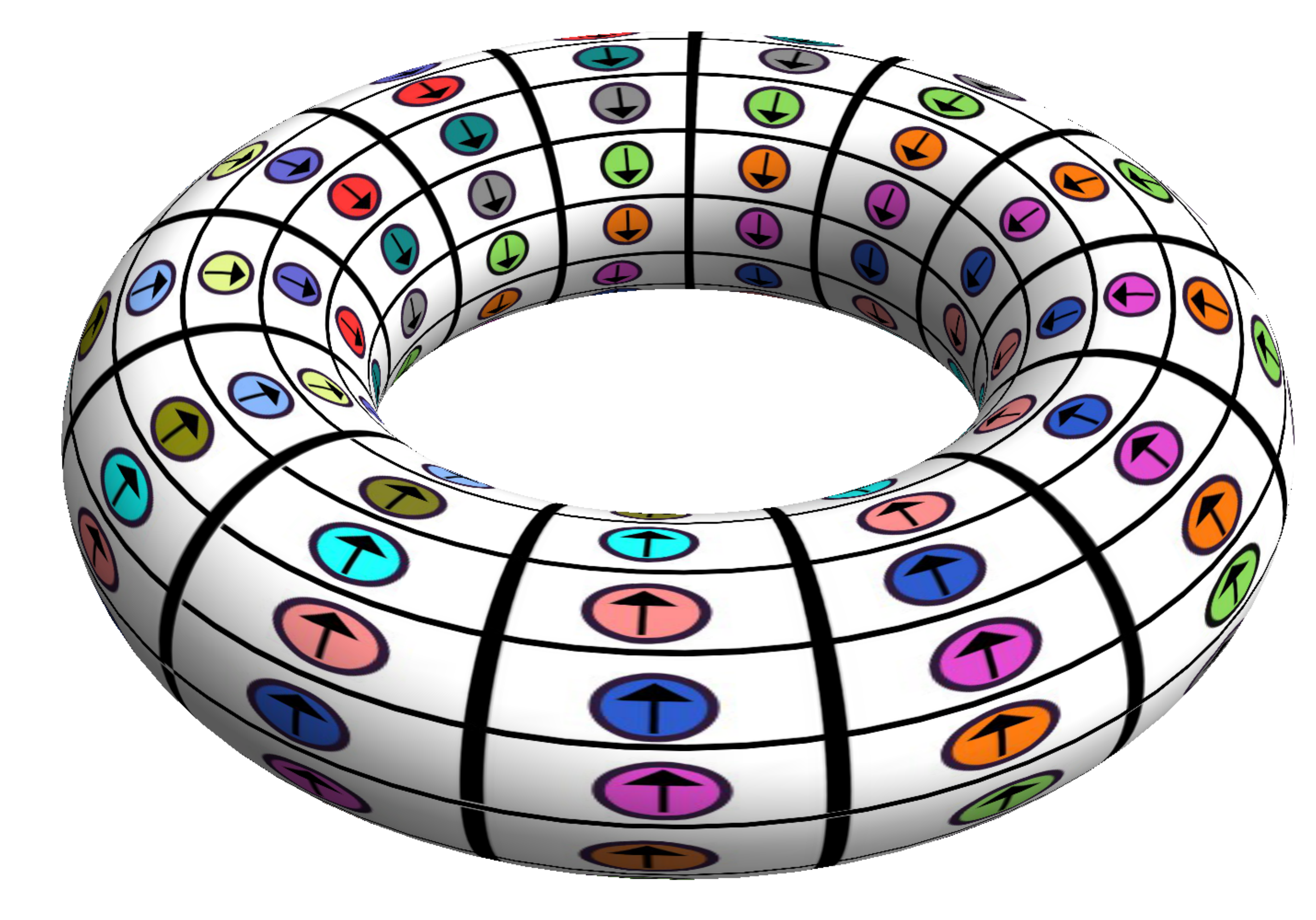}
    \put(-2,60){ \textbf c)}
    \end{overpic}
  \end{minipage}
  \begin{minipage}[b]{0.491\linewidth}
    \begin{overpic}[width=\linewidth]{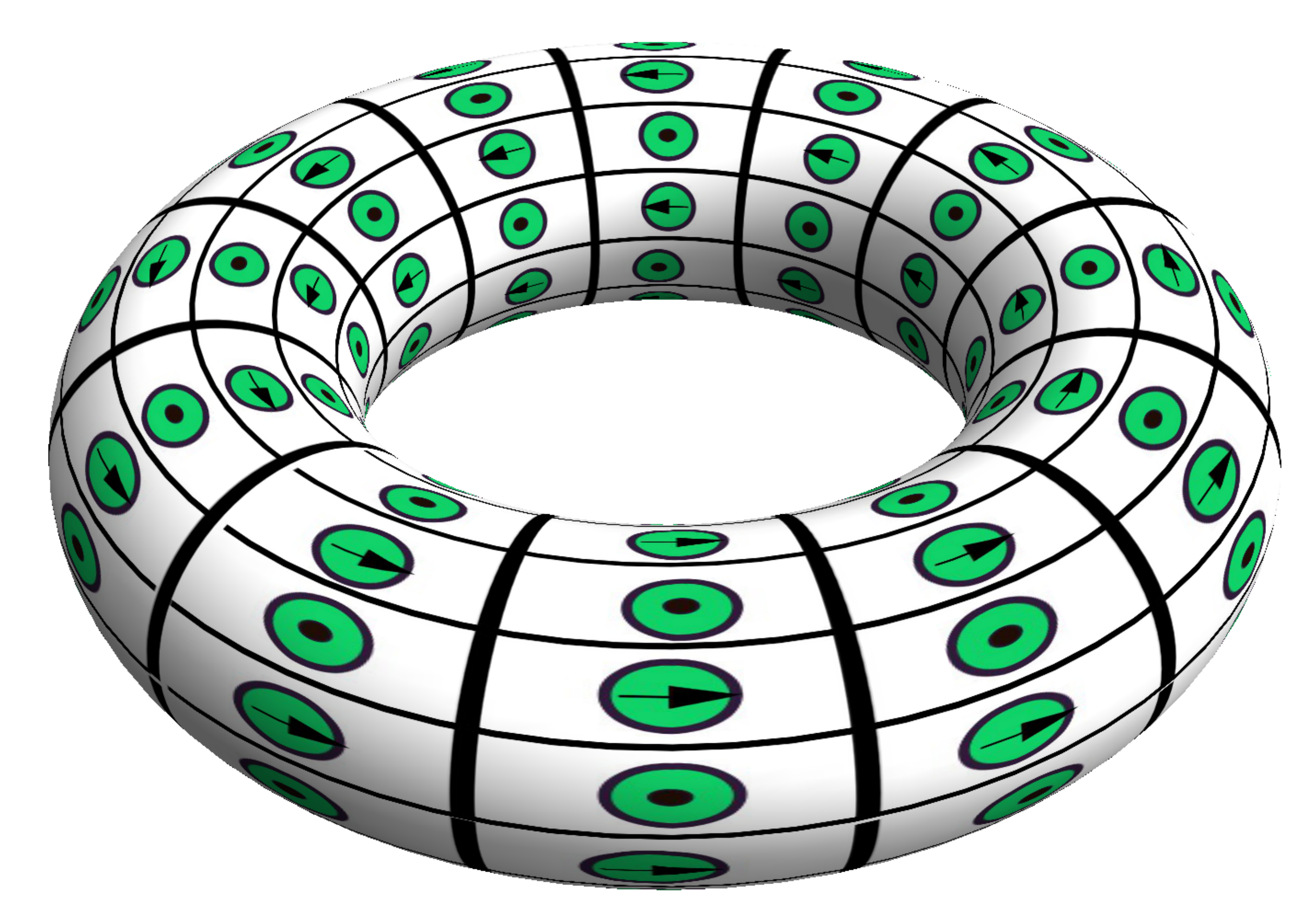}
    \put(-2,60){ \textbf d)}
    \end{overpic}
  \end{minipage}
  \caption{\label{FigW2}
  Caricature of the spin textures in the transverse Wen-plaquette model on the torus. (a) Depicts the topologically trivial spin texture, (b) depicts a topological state characteristic of the undriven model, and (c) depicts the dual spin configuration depicted in \ref{FigW1} (d). Finally, 
  (d) shows the driving-induced topological state (in the spin texture on the torus, the dots represent spins pointing to the positive $z$-direction).}
\end{figure}

To obtain the mapping, we observe that the operators $\hat{F}^{z}_{i,j}$ and $\hat{X}_{i,j}$ satisfy the commutation relations \cite{Wen1}
\begin{align}
      \label{WenOperatorComm}
            [\hat{F}^{z}_{a,b},\hat{X}_{c,d}] & = 2 \hat{F}^{z}_{a,b}\hat{X}_{c,d}(\delta_{a+1,b}+\delta_{a,b+1})\delta_{c,d},
            \nonumber \\
            [\hat{F}^{z}_{a,b},\hat{F}^{z}_{c,d}] & = [\hat{X}_{a,b},\hat{X}_{c,d}]=0
            .
\end{align}
Now, let us define a representation of this algebra by means of Pauli matrices $\tau^{x}_{i,j}$, $\tau^{y}_{i,j}$, and $\tau^{z}_{i,j}$ on the dual lattice
\begin{align}
       \label{MappingWenIsing} 
	     \hat{F}^{z}_{i^{*},j^{*}} & \mapsto \tau^{x}_{i^{*},j^{*}},
             \nonumber \\
	     \hat{X}_{i^{*},j^{*}} & \mapsto \tau^{z}_{i^{*},j^{*}}\tau^{z}_{i^{*},j^{*}+1}
       .
\end{align}
We parametrize the sites of the dual lattice by means of  $(i^{*},j^{*})=(i-j+1,j)$ following the notation of Ref. \cite{BrownVedral}. Under this duality transformation, the Hamiltonian Eq. \eqref{drivenWen} can be written in terms of $N$ decoupled 1-dimensional Ising chains of length $N$ \cite{Wen2,Wen1,BrownVedral}. In this paper $i^{*}$ denotes the chain index and $j^{*}$ the sites along the corresponding Ising chain.
In the dual representation, the TWPM Hamiltonian Eq. \eqref{drivenWen}  is written as
\begin{equation}
      \label{WenDecoupledIsingChains}
            \hat{H}(t)= \sum^{N}_{i^{*}=1}\hat{H}^{\tau}_{i^{*}}(t)
      ,
\end{equation}
where 
\begin{equation}
      \label{drivenWenIsingDual}
            \hat{H}^{\tau}_{i^{*}}(t) = -g(t)\sum_{j^{*}=1}^{N}\tau_{i^{*},j^{*}}^z\tau_{i^{*},j^{*}+1}^z-J\sum_{j^{*}=1}^{N}\tau_{i^{*},j^{*}}^x      
\end{equation}
is the Hamiltonian of the $i^{*}$-th Ising chain. 
The geometry of the decoupled Ising chains is shown in Fig.~\ref{FigW1} (d) and Fig.~\ref{FigW2} (c). 

Therefore, to understand the behavior of the TWPM under the effect of driving, we must discuss the quantum criticality in the driven Ising model Eq. \eqref{drivenWenIsingDual}. However, one can take advantage of the Kramers-Wannier self duality \cite{Peshel,Cirac}
\begin{align}
       \label{DualityIsing} 
	     \tau^{x}_{i^{*},j^{*}} & \mapsto \sigma^{z}_{i^{*},j^{*}}\sigma^{z}_{i^{*},j^{*}+1},
             \nonumber \\
	     \tau^{z}_{i^{*},j^{*}} & \mapsto \bigotimes_{r\leq j^{*}} \sigma^{x}_{i^{*},r} 
	,
\end{align}
and obtain a dual spin Hamiltonian
\begin{eqnarray}
\hspace{-0.4cm}
      \label{WendrivenIsing}
            \hat{H}^{\sigma}_{i^{*}}(t)&=& -g(t)\sum_{j^{*}=1}^{N}\sigma_{i^{*},j^{*}}^x-J\sum_{j^{*}=1}^N \sigma_{i^{*},j^{*}}^z \sigma_{i^{*},j^{*}+1}^z
      ,
\end{eqnarray}
which corresponds to an Ising model in a time-periodic transverse field \cite{Bastidas4}.
\subsection{TQPT from a spin-polarized state to topological ordered state in the undriven case \label{SubSectionI2}} 
In this section we review the more important aspects of the TQPT in the undriven TWPM following Refs. \cite{Wen2,Wen1}. 

The quantum Ising model exhibits a second-order QPT at $g_{0}=J$ from a paramagnetic phase into a ferromagnetic phase \cite{Sachdev}. However, to describe the TQPT in the TWPM one should define global order parameters as expectation values of nonlocal operators in the ground state of the system;
\begin{align}
      \label{GlobalOrderParameters}
            \Phi_1 &= \lim_{R\rightarrow \infty}\left\langle\bigotimes_{l=1}^{R} \hat{X}_{l,l}\right\rangle = 
            \lim_{R\rightarrow \infty}\left\langle\bigotimes_{j^{*}=1}^{R} \tau_{1,j^{*}}^{z}\tau_{1,j^{*}+1}^{z}\right\rangle 
            \\ \nonumber &= \lim_{R\rightarrow \infty}\left\langle\tau_{1,1}^{z}\tau_{1,R+1}^{z}\right\rangle,
            \nonumber \\
            \Phi_2 &= \lim_{R\rightarrow \infty}\left\langle\bigotimes_{l=1}^{R}\hat{F}^{z}_{l,l}\right\rangle =
            \lim_{R\rightarrow \infty}\left\langle \bigotimes_{j^{*}=1}^{R}\tau_{1,j^{*}}^{x}\right\rangle
            \\ \nonumber &= \lim_{R\rightarrow \infty}\left\langle\sigma_{1,1}^{z}\sigma_{1,R+1}^{z}\right\rangle
     .
\end{align}
In the topologically-ordered phase, $g_{0}<J$, the correlation function of $\sigma_{l}^z$ exhibits long-range order. As a consequence $\Phi_1 = 0$, and $\Phi_2 \sim \left(1-\left(\frac{g_{0}}{J}\right)^{2}\right)^{\frac{1}{4}}\neq 0$. This implies a macroscopic proliferation of $T_1$ and $T_2$ closed strings. In the spin-polarized phase (topologically trivial phase), $g_{0}>J$, a long-range order of the correlation function of $\tau_{l}^z$ indicates the condensation of open strings close to the TQPT, i.e., $\Phi_1 \sim \left(1-\left(\frac{J}{g_{0}}\right)^{2}\right)^{\frac{1}{4}}\neq 0$ and $\Phi_2 = 0$.

By using the duality between the Ising QPT and the TQPT in the TWPM, one we concludes
that for $g_{0}<J$, the paramagnetic phase of the $\tau$-spin model (ferromagnetic phase of the $\sigma$-spin model)  corresponds to the topologically-ordered (closed-string condensation) phase. Correspondingly, for $g_{0}>J$ the ferromagnetic phase of the $\tau$-spin model (paramagnetic phase of the $\sigma$-spin model) is dual to the spin-polarized phase of the TWPM (open string condensation) \cite{Wen2,Wen1,Deng}.
Figure \ref{FigW2} (a) depicts the spin-polarized phase of the TWPM on the torus and figure \ref{FigW2} (b) 
the topological phase. 

From  the last discussion, we conclude that the global order parameters $\Phi_1$ and $\Phi_2$ are dual \cite{Wen2,Wen1}. Therefore, to characterize the topological phase transition, only one of them is necessary. Consequently, in this paper we consider generalized order parameters to characterize the topological phases, and we will not discuss  the order parameter characterizing the spin-polarized phase.

\section{Floquet Topological Quantum Phase Transitions \label{SectionII}} 
As we mention in the last section, the topological phases of the undriven TWPM are closely related to the quantum phases of the Ising model. In the driven case, however, the Ising model Eq. \eqref{WendrivenIsing} exhibits a novel phase \cite{Bastidas4}. In this section we perform a description of the TWPM based on the RWA \cite{Nori}.
\subsection{The rotating wave approximation and the effective Hamiltonian approach \label{SubSectionII2}} 
Let us perform a unitary transformation of Hamiltonian Eq. \eqref{drivenWen} into a rotating frame via the unitary operator
\begin{align}
      \label{UnitaryWenRot}
            \hat{U}_{m}(t) &=
            \exp \left( \mathrm{i}\theta_{m}(t)\sum_{i,j}\hat{X}_{i,j}\right)=\bigotimes^{N}_{i^{*}=1} \hat{U}_{i^{*},m}(t)
            \nonumber \\
            &= \bigotimes^{N}_{i^{*}=1}\left[\exp \left( \mathrm{i}\theta_{m}(t)\sum_{j^{*}=1}^{N}\sigma_{i^{*},j^{*}}^x\right)\right]   
      ,
\end{align}
where $\theta_{m}(t)=m(\Omega/4) t+\frac{g_{1}}{\Omega}\sin\Omega t$. 
In the rotating frame, the Hamiltonian is given by
$\hat{H}_{m}(t)=\hat{U}^\dagger_{m}(t) \hat{\mathcal{H}} \hat{U}_{m}(t)$, where $\hat{\mathcal{H}}=\hat{H}(t)-\mathrm{i} \frac{\partial}{\partial t}$ is the Floquet Hamiltonian \cite{Shirley}. The explicit form of this operator is given by
\begin{align}
      \label{HamiltonianWenRotFrameSI}
	    \hat{H}_{m}(t)&= -\Delta_{m}\sum_{i,j}\hat{X}_{i,j} 
	    \nonumber \\&	    
	    -\frac{J}{2}\left\{1+\cos[4\theta_{m}(t)]\right\}\sum_{i,j}\hat{X}_{i,j} \hat{Y}_{i+1,j}
            \hat{X}_{i+1,j+1} \hat{Y}_{i,j+1}
            \nonumber \\&
            -\frac{J}{2}\left\{1-\cos[4\theta_{m}(t)]\right\}\sum_{i,j}\hat{X}_{i,j} \hat{Z}_{i+1,j}
            \hat{X}_{i+1,j+1} \hat{Z}_{i,j+1}
	    \nonumber \\&
	    +\frac{J}{2}\sin[4\theta_{m}(t)]\sum_{i,j} \hat{X}_{i,j} \hat{Y}_{i+1,j} \hat{X}_{i+1,j+1}\hat{Z}_{i,j+1}
	    \nonumber \\&
	    +\frac{J}{2}\sin[4\theta_{m}(t)]\sum_{i,j}\hat{X}_{i,j} \hat{Z}_{i+1,j}\hat{X}_{i+1,j+1} \hat{Y}_{i,j+1}
      ,
\end{align}
where $\Delta_{m}=g_{0}-m (\Omega/4)$ is the detuning from the $m$-photon resonance \cite{Shirley}. 

The Hamiltonian Eq. \eqref{HamiltonianWenRotFrameSI} can be expanded in Fourier series
\begin{align}
      \label{WenHamiltonianFourier}
             \hat{H}_{m}(t)&=\sum_{n=-\infty}^{\infty} \hat{h}^{(m)}_{n}\exp{(\mathrm{i} n \Omega t)}
             \nonumber \\&
             =\sum_{n=-\infty}^{\infty}\sum^{N}_{i^{*}=1}\hat{h}^{(\sigma,m)}_{i^{*},n}\exp{(\mathrm{i} n \Omega t)}
   ,
\end{align}
where $\hat{h}^{(m)}_{n}$ and $\hat{h}^{(\sigma,m)}_{i^{*},n}$ are time-independent operators.

We define our effective Hamiltonian as an average of the Hamiltonian in the rotating frame $ \hat{H}_{m}(t)$
over a period $T=2\pi/\Omega$ of the driving
\begin{equation}
      \label{WenIsingRWA}
             \hat{h}^{(m)}_{0}=\int^{T}_{0}\frac{dt}{T}\hat{H}_{m}(t)=\sum^{N}_{i^{*}=1}\hat{h}^{(\sigma,m)}_{i^{*},0}
      ,
\end{equation}
where $\hat{h}^{(\sigma,m)}_{i^{*},0}$ is the effective Hamiltonian given in Eq. \eqref{mResonaceIsingXY}.
Under the conditions 
\begin{equation}
      \label{ValidityRWAIsing}
            \Delta_{m},J\mathcal{J}_{m}\left(\frac{4g_{1}}{\Omega}\right) \ll\Omega
\end{equation}
the fast oscillating terms in the Hamiltonian Eq. \eqref{HamiltonianWenRotFrameSI} can be safely neglected and the effective Hamiltonian $\hat{h}^{(m)}_{0}$ governs the dynamics in the rotating frame. In the last equation, $\mathcal{J}_{l}(z)$ is the  $l$th-order Bessel function \cite{Abramowitz}.

The time-independent effective Hamiltonian in the real lattice reads
\begin{align}
      \label{mResonaceWen}
            \hat{h}^{(m)}_{0} &= -\Delta_{m}\sum_{i,j}\hat{X}_{i,j}
            -J_{z}^{(m)}\sum_{i,j}\hat{X}_{i,j} \hat{Y}_{i+1,j}
            \hat{X}_{i+1,j+1} \hat{Y}_{i,j+1}
             \nonumber \\&
            -J_{y}^{(m)}\sum_{i,j}\hat{X}_{i,j} \hat{Z}_{i+1,j}
            \hat{X}_{i+1,j+1} \hat{Z}_{i,j+1}
             \nonumber \\
             &=-\Delta_{m}\sum_{i,j}\hat{X}_{i,j}-J_{z}^{(m)}\sum_{i,j}\hat{F}^{z}_{i,j}-J_{y}^{(m)}\sum_{i,j}\hat{F}^{y}_{i,j}
      ,
\end{align}
where the parameters $J^{(m)}_z=\frac{J}{2}[1+(-1)^m\mathcal{J}_{m}(\frac{4 g_{1}}{\Omega})]$ and $J^{(m)}_y=\frac{J}{2}[1-(-1)^m\mathcal{J}_{m}(\frac{4 g_{1}}{\Omega})]$ are effective anisotropies in the rotating frame \cite{Bastidas4}. 

Besides the plaquette operator $\hat{F}^{z}_{i,j}$ defined in Eq. \eqref{Plaquetteoperator}, the effective Hamiltonian Eq. \eqref{mResonaceWen} also contains a driving-induced plaquette operator
\begin{eqnarray}
      \label{NewPlaquetteoperator}
            \hat{F}^{y}_{i,j} = \hat{X}_{i,j} \hat{Z}_{i+1,j}
            \hat{X}_{i+1,j+1} \hat{Z}_{i,j+1}
       .
\end{eqnarray}
This new term is absent in the original undriven TWPM and corresponds to an effective interaction originated under nonequilibrium conditions \cite{Bastidas3}.  

\subsection{AC-driven Quantum Phase transitions in the Ising model \label{SubSectionII1}} 
Our aim in this section is to describe the phase diagram of the driven TWPM Hamiltonian Eq. \eqref{drivenWen} in terms of the quantum phases of the driven Ising model Eq. \eqref{WendrivenIsing} studied in Ref. \cite{Bastidas4}. 

The criticality in the rotating frame is described by means of the effective Hamiltonian
\begin{eqnarray}
      \label{mResonaceIsingXY}
             \hat{h}^{(\sigma,m)}_{i^{*},0} &=& -\Delta_{m}\sum_{j^{*}=1}^{N} \sigma_{i^{*},j^{*}}^x-J^{(m)}_z\sum_{j^{*}=1}^N  \sigma_{i^{*},j^{*}}^z\sigma_{i^{*},j^{*}+1}^z
             \nonumber \\&&
             -J^{(m)}_y\sum_{j^{*}=1}^N \sigma_{i^{*},j^{*}}^y
             \sigma_{i^{*},j^{*}+1}^y
      ,
\end{eqnarray}
which is derived explicitly in Appendix \ref{AppendixA}.

Similarly to the Ising model \cite{Sachdev}, the Hamiltonian Eq. \eqref{mResonaceIsingXY} can be written in terms of Jordan-Wigner fermionic operators
$\hat{c}_{k}$ and $\hat{c}^{\dagger}_{k}$ as follows
\begin{align}
      \label{FourierIsingEffective}
             \hat{h}^{(\sigma,m)}_{i^{*},0}&= \sum_{k \geq 0}\left[(2\Delta_{m}-\omega_k)(\hat{c}_k^{\dagger}\hat{c}_k+\hat{c}_{-k}^{\dagger}\hat{c}_{-k})-2\Delta_{m}\hat{\mathbbm{1}}_{k}\right]  
             \nonumber \\
             & +\sum_{k \geq 0}(-1)^{m} \mathcal{J}_{m}\left(\frac{4 g_{1}}{\Omega}\right)f_k(\hat{c}_k^{\dagger}\hat{c}_{-k}^{\dagger}+\hat{c}_{-k} \hat{c}_{k})            
      ,
\end{align}
where $\omega_k=2J \cos k$ and $f_{k}=2J\sin k$ \cite{Bastidas4}.

The Hamiltonian Eq. \eqref{FourierIsingEffective} can be written as a Hamiltonian for free fermions
\begin{eqnarray}
      \label{DiagonalizedXYmodel}
             \hat{h}^{(\sigma,m)}_{i^{*},0} =\sum_{k \geq 0} \varepsilon_{k,m}\left(\hat{\gamma}_k^{\dagger}\hat{\gamma}_k+\hat{\gamma}_{-k}^{\dagger}\hat{\gamma}_{-k}-\hat{\mathbbm{1}}_{k}\right)
\end{eqnarray}
with dispersion relation 
\begin{equation}
      \label{QuasiEnergyExcitation}
            \varepsilon_{k,m}=2\sqrt{\left(\Delta_{m}-J\cos k\right)^2+\left[J\mathcal{J}_{m}\left(\frac{4 g_{1}}{\Omega}\right)\sin k\right]^2}.
\end{equation}
Therefore, when the gap between the positive and negative energies closes, the effective Hamiltonian exhibits signatures of criticality in the rotating frame. Based on the results of Ref. \cite{Bastidas4}, we find that the effective Hamiltonian Eq. \eqref{mResonaceIsingXY} exhibits effective ac-driven Ising-like QPTs along the critical lines
$|\Delta_{m}| = J$. Via the duality transformations Eqs. \eqref{MappingWenIsing} and \eqref{DualityIsing}, the Ising-like QPT in the Ising model corresponds to a transition between the trivial and the topological phases in the driven TWPM. 

The Hamiltonian Eq. \eqref{mResonaceIsingXY} also exhibit ac-driven anisotropic QPTs when the effective anisotropy  $\gamma^{(m)}=\mathcal{J}_{m}\left(\frac{4 g_{1}}{\Omega}\right)$ vanishes, under the condition $|\Delta_{m}| < J$. The anisotropic transitions occur between two ferromagnetic phases. For $J^{(m)}_{z}> J^{(m)}_{y}$, the system exhibits long-range magnetic order along $z$ direction (ferromagnetic phase FM$Z$), while for for $J^{(m)}_{z}< J^{(m)}_{y}$, the dual Ising model Eq. \eqref{mResonaceIsingXY} is ferromagnetically ordered along the $y$ direction (ferromagnetic phase FM$Y$) \cite{Bastidas4}. Interestingly, the anisotropic transition in the dual Ising model corresponds to an ac-driven anisotropic transition between topological phases of the TWPM.
\begin{figure}
\centering
\includegraphics[width=0.95\linewidth]{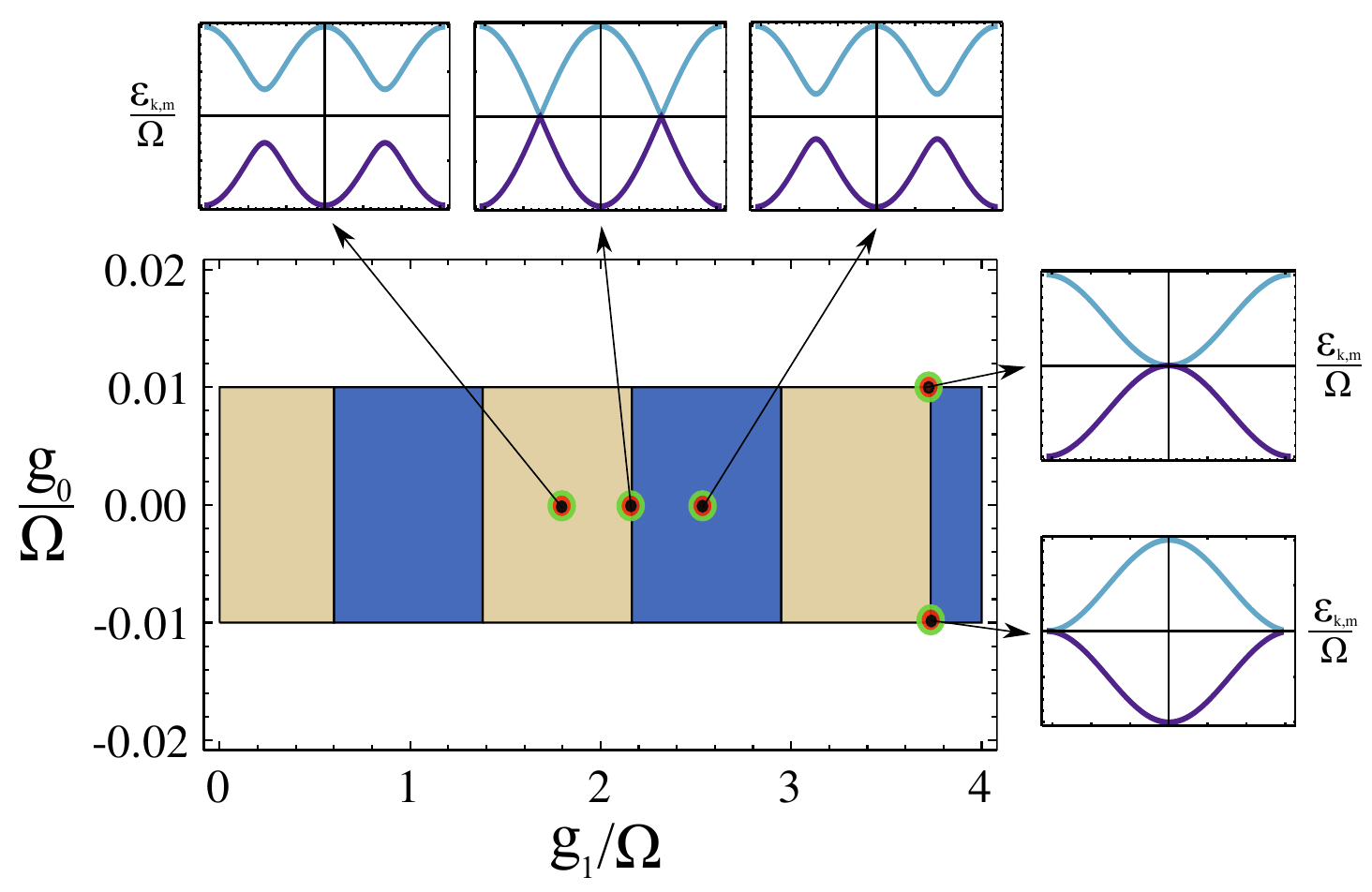}
\caption{
  \label{FigI2}
  Quantum phase diagram of the Wen-plaquette model in a time-dependent transverse field $g(t)=g_0+g_1\cos\Omega t$. The figure depicts the quasienergy dispersion $\pm \varepsilon_{k,m}$ of the dual Ising model for parameters at the Ising-like critical lines, in the ferromagnetic phases, and along an anisotropic critical line $\gamma^{(m)}=0$. For this plot we consider $J/\Omega=0.01$ and $m=0$.
}
\end{figure}
Figure \ref{FigI2} depicts the phase diagram for the ac-driven TQPT in TWPM for $m=0$. The white zones in the phase diagram correspond to the topologically trivial phase (effective paramagnetic phase of the dual Ising model Eq. \eqref{WendrivenIsing}) and are defined by the inequality 
$J<|\Delta_{m}|<|\Delta_{\text{max}}|$, for $m=0$, where $\Delta_{\text{max}}$ denotes the maximum detuning for which the RWA is still valid. In Fig.~\ref{FigW4} (a) we plot the the effective asymmetries $J^{(m)}_{z}$ and $J^{(m)}_{y}$ as a function of the driving amplitude.

In the undriven TWPM the topological phases are characterized by nonlocal order parameters. However, as one can see after the last discussion, there is necessary
to quantify the entities arising under the effect of driving. In the next section we define generalized topological order parameters.
 
\section{Cycle-averaged Topological Order Parameters \label{SectionIII}} 
We now calculate global cycle-averaged expectation values of nonlocal operators. In particular, we focus on the expectation values in the negative-quasienergy Floquet state of the driven TWPM 
\begin{equation}
      \label{NegativeQEWen}
            \left|\Psi^{(-)}_{W,m},t\right\rangle=\bigotimes_{i^{*}=1}^{N}\left|\Psi^{(-)}_{i^{*},m},t\right\rangle
      ,
\end{equation}
where
\begin{align}
      \label{GeneralIsingFloquetstate}
           \left|\Psi^{(\pm)}_{i^{*},m},t\right\rangle &=\bigotimes_{k \geq 0}\left|\Psi^{(\pm)}_{k,m},t\right\rangle           
           = \bigotimes_{k \geq 0}e^{-\mathrm{i}\varepsilon_{k,m}^{(\pm)}t}\left|\Phi^{(\pm)}_{k,m},t\right\rangle
           \\ \nonumber &
            = \exp(-\mathrm{i} N E^{(\pm)}_{m}t)\bigotimes_{k \geq 0}\left|\Phi^{(\pm)}_{k,m},t\right\rangle
\end{align}
is a Floquet state of the $i^{*}$-th Ising chain.
In the last expression, $E^{(\pm)}_{m}=\pm\int_0^{\pi}\frac{dk}{2\pi}\varepsilon_{k,m}$ is the total quasienergy and $\varepsilon_{k,m}$ is the quasienergy 
dispersion defined in Eq. \eqref{QuasiEnergyExcitation}. Furthermore, as we describe in Appendix \ref{AppendixA}, the Floquet $k$-eigenmodes of the $i^{*}$-th Ising
are obtained by applying an unitary transformation back into the laboratory frame 
$|\Phi^{(\pm)}_{k,m},t\rangle=\hat{V}_{k,m}(t)|\chi^{(\pm)}_{k,m}\rangle$, with
\begin{align}
      \label{IsingEffecEigenVectors}
      \left|\chi^{(+)}_{k,m}\right\rangle &= \cos\phi_{k,m}|1_{-k},1_{k}\rangle-\sin\phi_{k,m}|0_{-k},0_{k}\rangle,
      \nonumber \\
      \left|\chi^{(-)}_{k,m}\right\rangle &= \sin\phi_{k,m}|1_{-k},1_{k}\rangle+\cos\phi_{k,m}|0_{-k},0_{k}\rangle
      ,
\end{align}
satisfying $\hat{h}^{(m)}_{i^{*},0}|\chi^{(\pm)}_{k,m}\rangle=\pm\varepsilon_{k,m}|\chi^{(\pm)}_{k,m}\rangle$. We have used the basis of doubly occupied $|1_{-k},1_{k}\rangle$ and unoccupied $|0_{-k},0_{k}\rangle$ states of $\pm k$ fermions. Furthermore, we consider only the subspace with vanishing total momentum \cite{Sachdev}.

The Bogoliubov angle $\phi_{k,m}$ is determined by the relation 
\begin{equation}
      \label{BogoliuvobAngle}
            \tan(2\phi_{k,m})=\frac{-(-1)^{m}\mathcal{J}_{m}\left(\frac{4 g_{1}}{\Omega}\right)f_k}{2\Delta_{m}-\omega_k}.
\end{equation}

In the limit $g_{0}\gg J$, the negative-quasienergy Floquet state Eq. \eqref{NegativeQEWen} corresponds to a topologically trivial spin
texture in the absence of driving [see Fig. \ref{FigW2} (a)]. In the dual picture of the $\sigma$-spins, such a state is a tensor product of stationary states of the Hamiltonian Eq. \eqref{WendrivenIsing} with all the spins polarized along the $x$ axis 
\begin{equation}
      \label{InitialParamagneticStateWen}
            \left|\Psi^{(-)}_{W,m},t\right\rangle=\exp\left(-\mathrm{i}N^{2}E_G^{(-)}t\right)\bigotimes_{i^{*}=1}^{N}\bigotimes_{k\geq 0}|0_{-k},0_{k}\rangle
      .
\end{equation}
In the last expression
\begin{equation}
      \label{IsingGroundEnergy}
            E_G^{(-)}=
            -\frac{2}{\pi}\sqrt{g^2_{0}+J^2}E\left[\frac{4g_{0}J}{(g_0+J)^{2}}\right]
\end{equation}
is the ground-state energy of the undriven Ising model \cite{Sachdev}, where $E[z]$ is the complete elliptic integral of the second kind \cite{Abramowitz}. 

\subsection{Topological order from long-range magnetic order in the dual Ising model \label{SubSectionIII1}} 
We focus here on the calculation of cycle-averaged global order parameters. The generalized closed-string order parameters are defined as products of plaquette operators along the diagonal string of the real lattice.

By using the spin dualities Eq. \eqref{MappingWenIsing} and Eq. \eqref{DualityIsing} we define 
\begin{align}
      \label{CylceAvGlobalOrderParameters} 
            \overline{\Phi}_2 &= \lim_{R\rightarrow \infty}
            \frac{1}{T}\int_{0}^{T}\left\langle\bigotimes_{l=1}^{R}\hat{F}^{z}_{l,l}\right\rangle \ dt 
            \nonumber \\
            &=\lim_{R\rightarrow \infty}\frac{1}{T}\int_{0}^{T}\left\langle\sigma_{1,1}^{z}\sigma_{1,1+R}^{z}\right\rangle \ dt ,
            \nonumber \\
            \overline{\Phi}_3 &= \lim_{R\rightarrow \infty}
            \frac{1}{T}\int_{0}^{T}\left\langle\bigotimes_{l=1}^{R}\hat{F}^{y}_{l,l}\right\rangle \ dt 
            \nonumber \\
            &=\lim_{R\rightarrow \infty}\frac{1}{T}\int_{0}^{T}\left\langle\sigma_{1,1}^{y}\sigma_{1,1+R}^{y}\right\rangle \ dt
     .
\end{align}
Our definition of $\overline{\Phi}_2$ is a generalization to driven quantum systems of the global order parameters defined in equilibrium \cite{Wen2,Wen1}. However, the existence of $\overline{\Phi}_3 $ has no analogue in equilibrium and a nonvanishing value of it would give rise to a topological configuration induced entirely by the external control.
\begin{figure}
\vspace{0.4cm}
  \begin{minipage}[b]{0.81\linewidth}
  \hspace{-0.7cm}
    \begin{overpic}[width=\linewidth]{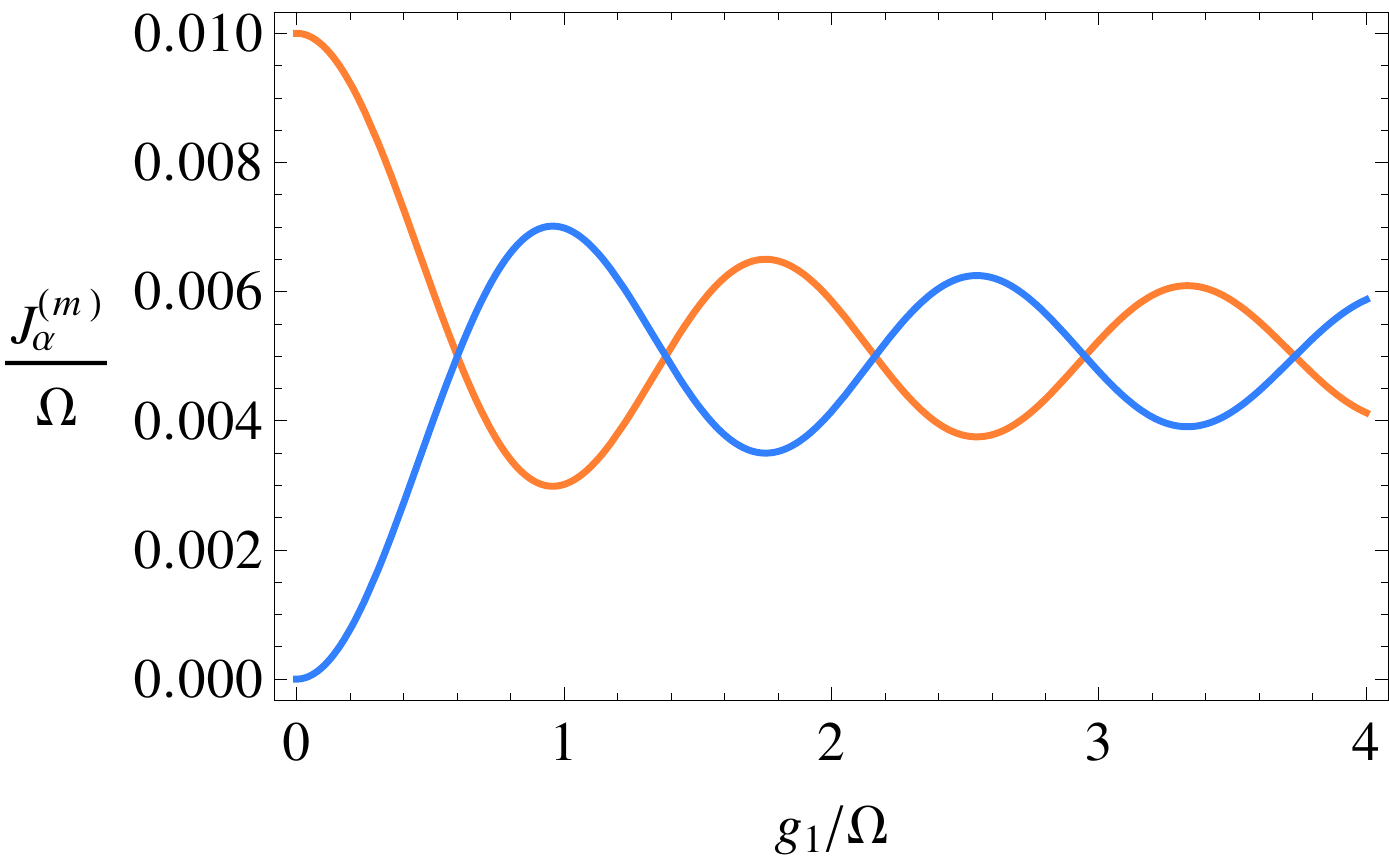}
    \put(0.5,60){ \textbf a)}
    \end{overpic}
  \end{minipage}
  
  \begin{minipage}[b]{0.74\linewidth}
  \hspace{2.4cm}
    \begin{overpic}[width=\linewidth]{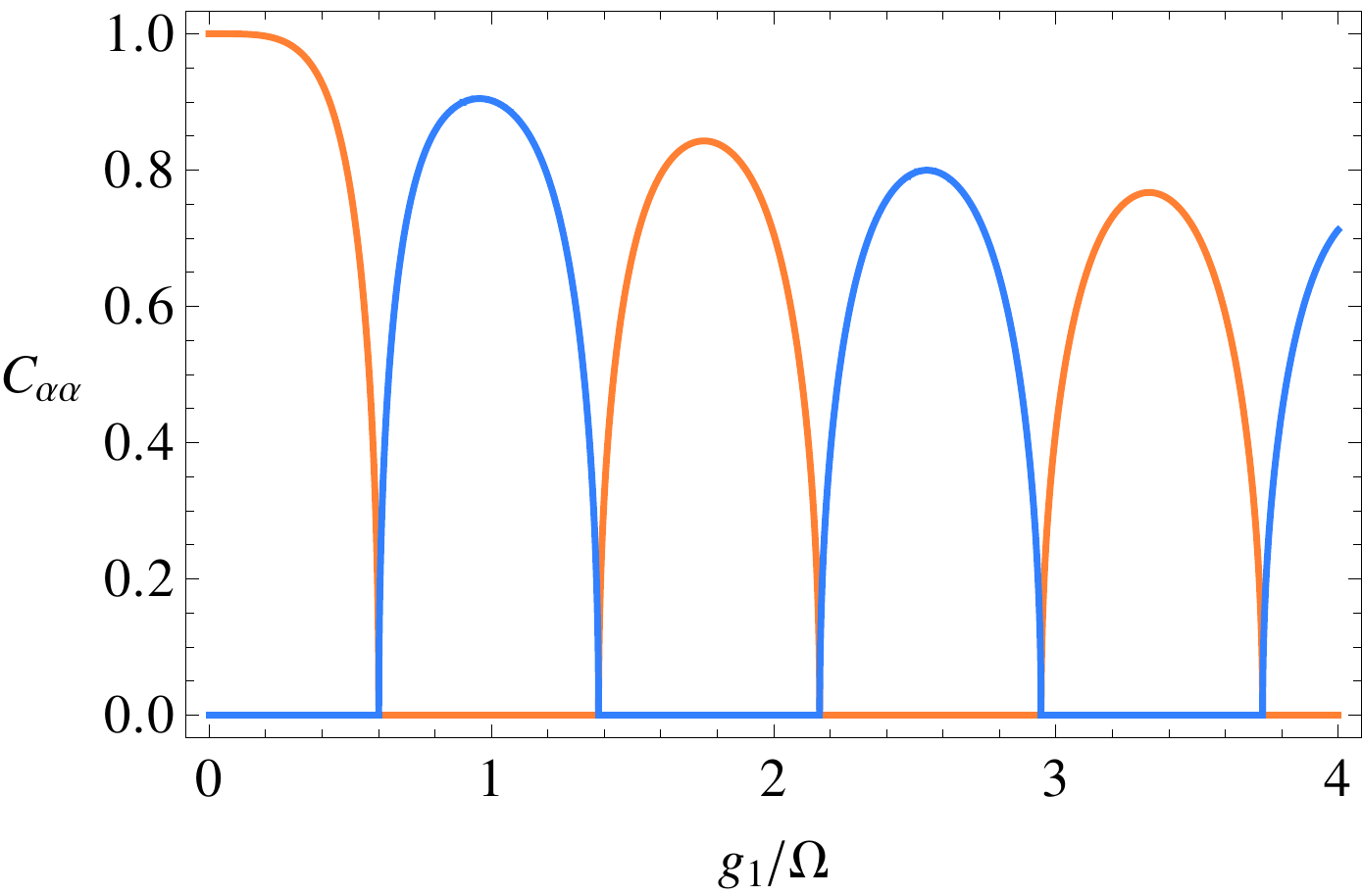}
    \put(-5,60){ \textbf b)}
    \end{overpic}
  \end{minipage}
  \hspace{+1.9cm}
   \begin{minipage}[b]{0.715\linewidth}
   \hspace{+1.9cm}
    \begin{overpic}[width=\linewidth]{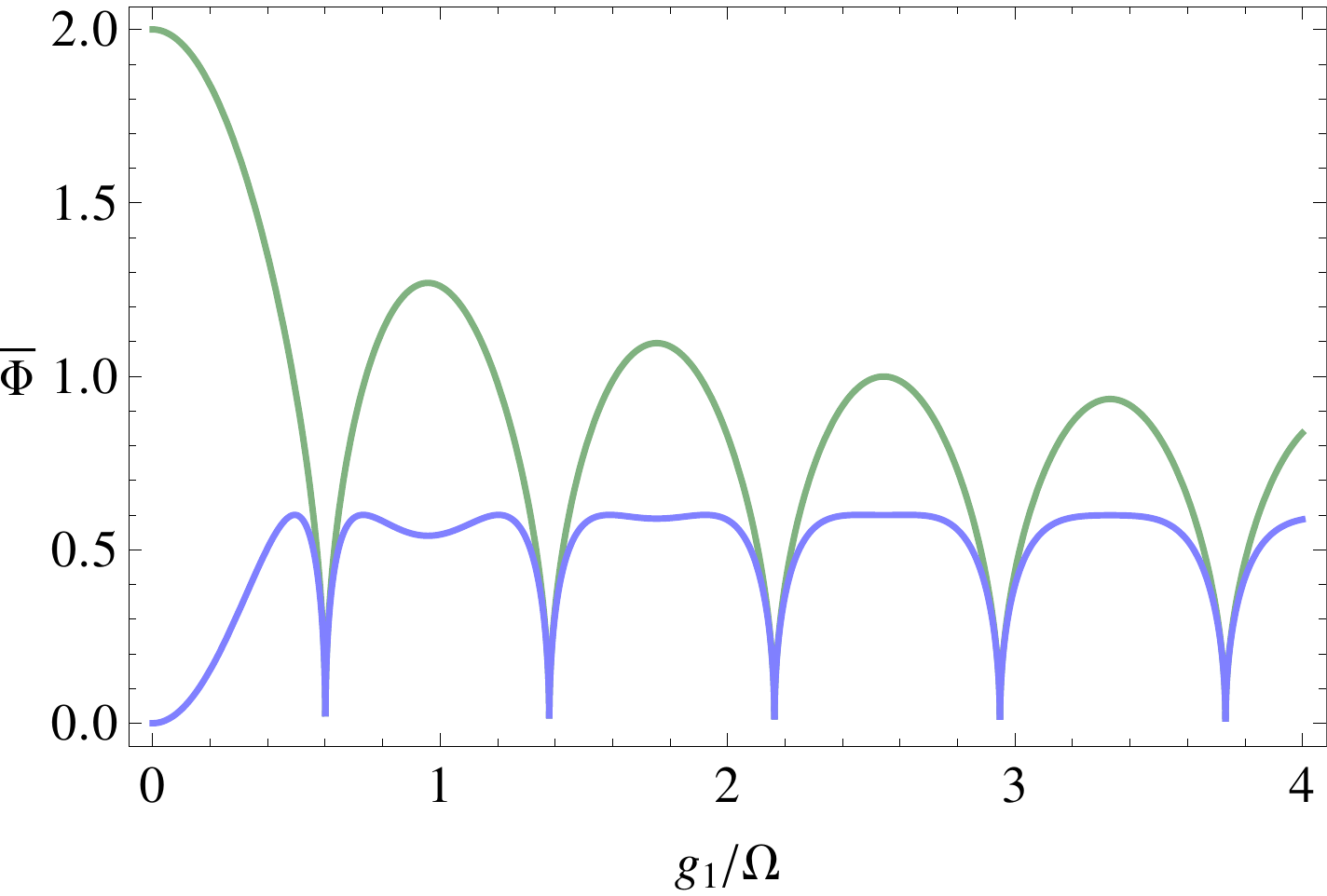}
    \put(-6,63){ \textbf c)}
    \end{overpic}
  \end{minipage}
  \caption{\label{FigW4}
  (a) Depicts the effective anisotropies. The orange curve corresponds to $J^{m}_{z}$
  and the blue curve to $J^{m}_{y}$. (b) Shows the nonlocal order parameters in the rotating frame. The orange curve denotes the order parameter $C_{zz}$ and the blue curve depicts $C_{yy}$. (c) Shows the cycle-averaged order parameters. The green curve depicts $\overline{\Phi}_2$, and the violet curve represents $\overline{\Phi}_3$. We consider $\alpha \in \{y,z\}$, and the parameters $J/\Omega=0.01$ and $m=0$.}
\end{figure}

Interestingly, the cycle-averaged order parameters correspond to cycle-averaged correlation functions of the dual 1-dimensional driven Ising model Eq. \eqref{WendrivenIsing}. Such correlation functions characterize the long-range magnetic order under the effect of external driving.
One can calculate the cycle-averaged nonequilibrium spin correlation functions in terms of the well-known results for the $XY$ model \cite{Mattis, Barouch,SadchevXY}.

For $\alpha \in \{y,z\}$, let us define nonlocal order parameters in the rotating frame
$C_{\alpha\alpha}=\lim_{R\rightarrow \infty}\rho_{\alpha \alpha}(R)$, where
\begin{equation}
      \rho_{\alpha \alpha}(R)=\langle\chi^{(-)}_{m}|\sigma_{1}^{\alpha}\sigma_{1+R}^{\alpha}|\chi^{(-)}_{m}\rangle 
\end{equation}
corresponds to a spin correlation function calculated in the rotating frame (see Appendix \ref{AppendixB}), where
\begin{equation}
      \label{Rot}
            \left|\chi^{(-)}_{m}\right\rangle = \bigotimes_{k\geq 0}\left|\chi^{(-)}_{k,m}\right\rangle
      ,
\end{equation}
and $|\chi^{(-)}_{k,m}\rangle$ is given by Eq. \eqref{IsingEffecEigenVectors}.

We can explicitly write (as we show in Appendix \ref{AppendixB}) the topological order parameters as follows:
\begin{align}
      \label{OrderTopYZ}
            \overline{\Phi}_2 &= \frac{2}{J}[J_{z}^{(m)}C_{zz}+J_{y}^{(m)}C_{yy}] 
            \nonumber \\
            \overline{\Phi}_3 &= \frac{2}{J}[J_{z}^{(m)}C_{yy}+J_{y}^{(m)}C_{zz}]
            ,
\end{align}
where $J_{z}^{(m)}$, $J_{y}^{(m)}$ are effective anisotropies as they appear in Hamiltonian Eq. \eqref{mResonaceIsingXY}. 

We find that $\overline{\Phi}_2=\overline{\Phi}_3=0$ in the topologically trivial phase $|\Delta_{m}|>J$.
For $|\Delta^{m}|<J$ the long-range correlation functions are given by
\begin{align}
      \label{RotatingFramCorrelations}
C_{zz} &=\left\{%
\begin{array}{cc}
\frac{2(|\gamma^{(m)}|)^{1/2}}{1+|\gamma^{(m)}|}\left[1-\left(\frac{\Delta_{m}}{J}\right)^{2}\right]^{1/4} & \gamma^{(m)}>0\\
0 & \gamma^{(m)}<0
\end{array}
\right. ,
\nonumber \\
C_{yy} &=\left\{%
\begin{array}{cc}
0 & \gamma^{(m)}>0 \\
\frac{2(|\gamma^{(m)}|)^{1/2}}{1+|\gamma^{(m)}|}\left[1-\left(\frac{\Delta_{m}}{J}\right)^{2}\right]^{1/4} & \gamma^{(m)}<0
\end{array}
\right.
,
\end{align}
where $\gamma^{(m)}=(-1)^{m}\mathcal{J}_{m}\left(\frac{4g_{1}}{\Omega}\right)$ is the anisotropy parameter. Fig. \ref{FigW4} (c) depicts the nonlocal order parameters in the rotating frame as a function of the driving amplitude. 

As a limiting case, our description recovers the physics of the undriven model $(g_{1}=0)$ in the particular case $m=0$, where $\Delta_{0}=g_{0}$, $\gamma^{(0)}=1$, $J_{z}^{(0)}=J$, and $J_{y}^{(0)}=0$. Therefore, we obtain that the nonlocal order parameters satisfy
$\overline{\Phi}_2=\Phi_2$ and $\overline{\Phi}_3=0$. 

We plot the dependence of the order parameters along the ladder of topological phases (ferromagnetic phases FM$Z$ and FM$Y$ from the dual Ising model depicted in Fig.\ref{FigI2}) in Fig. \ref{FigW4} (c). Our results show that even for arbitrarily weak driving amplitude $g_1$, it is possible to drive the system from the trivial phase into the topological phase. Surprisingly, in the case $m>0$, the static parameters 
are far from the critical point of the undriven system, i.e., $g^{c}_0=J$. For example, in the $m>0$ case, exact resonance $\Delta_{m}=0$ implies to work in the regime where the undriven system is in the trivial phase $g_0=m\Omega/2 \gg J$. The strong driving regime allows to explore a new aspect of our approach: by tuning the amplitude of the external control, it is possible to bring the system into a new topological phase. The phase diagram
depicted in Fig. \ref{FigI2} shows rich patterns of multicritical points with no analogue in equilibrium when $\rvert \Delta_{m}=J\rvert$ and $\gamma^{(m)}=0$.   

\section{Conclusions\label{SectionIV}}
We have discussed the nonequilibrium TQPT in the Wen-plaquette model in a time-dependent transverse field. Similarly to the undriven model, a set of highly nonlocal spin-duality transformations allows a description of the driven TWPM in terms of the Ising model physics. The topological character of the TWPM requires a description of the TQPT in terms of nonlocal order parameters, which correspond to long-range correlation functions for a driven Ising model in the dual picture. We have introduced generalized ``string''-like topological order parameters by considering cycle-averaged expectation values of string operators in a Floquet state.

In the case of conventional nonequilibrium QPTs, previous works 
\cite{Bastidas3,Bastidas4,BastidasLMG,HolthausQPT,Vedral,Creffield} show that even under the effect of driving, the quantum states of matter correspond to symmetry broken phases in the thermodynamic limit, i.e., the superradiant state does not conserve parity symmetry as in Ref. \cite{Bastidas3}. In contrast, a TQPT corresponds to a change of phase without symmetry breaking. Similar to Refs. \cite{ZollerTopo,Lindner}, our approach reveals a possibility to induce topological configurations of the system by driving the topologically trivial phase. We show, however, that the monochromatic driving not only renormalizes the critical point, but generates an additional topological phase.

Our methodology is potentially interesting in the context of quantum simulation with cold atoms. A promising experimental realization of the driven Wen-plaquette model could be achievable by using a Rydberg atom quantum simulator \cite{ZollerTC}. In this setup, to simulate the system it is necessary to construct a quantum circuit consisting of nonlocal gates that encode the interactions of the effective model.

\begin{acknowledgments}
 V. M. Bastidas acknowledges fruitful discussions with H. Gomez Zu\~{n}iga and K. P. Schmidt. The authors gratefully acknowledge discussions with M. Hayn, I. Lesanovsky, C. Nietner, P. Strasberg, P. Strack and M. Vogl, and financial support from the DAAD, DFG Grants BR $1528/7-1$, $1528/8-1$, SFB $910$, GRK $1558$, and SCHA $1646/2-1$.
 This work was also supported through the Grants No. MAT$2011-24331$ and ITN, No. $234970$ (EU).
\end{acknowledgments}
%

%
\appendix
%
\section{The rotating wave approximation and the effective Hamiltonian of the Ac-driven Ising model \label{AppendixA}}
To obtain the effective Hamiltonian Eq. \eqref{mResonaceIsingXY}, we perform a description of the system based on the rotating wave approximation \cite{Bastidas4,Nori}. Let us perform a unitary transformation of Hamiltonian Eq. \eqref{WendrivenIsing} into a convenient rotating frame via the unitary operator
\begin{align}
      \label{AppUnitaryIsingRot}
            \hat{U}_{i^{*},m}(t) &=
            \exp \left( i\theta_{m}(t)\sum_{j^{*}=1}^{N}\sigma_{i^{*},j^{*}}^x\right)=\bigotimes_{k>0} \hat{V}_{k,m}(t)
            \nonumber \\
            =& \bigotimes_{k>0} \exp \left[-2i\theta_{m}(t)(c_k^{\dagger}c_k+c_{-k}^{\dagger}c_{-k}-\hat{\mathbbm{1}}_{k})\right]
      ,
\end{align}
where $\theta_{m}(t)=m(\Omega/4) t+\frac{g_{1}}{\Omega}\sin\Omega t$, and $c_k$ are Jordan-Wigner fermions.

In the rotating frame the dynamics is governed by the new Hamiltonian 
$\hat{H}_{i^{*},m}^{\sigma}(t)=[\hat{U}_{i^{*},m}(t) ]^\dagger \hat{\mathcal{H}_{i^{*}}} \hat{U}_{i^{*},m}(t) $. The operator $\hat{\mathcal{H}_{i^{*}}}=\hat{H}^{\sigma}_{i^{*}}(t)-i \frac{\partial}{\partial t}$ is the Floquet Hamiltonian \cite{Shirley} and the Hamiltonian $\hat{H}^{\sigma}_{i^{*}}(t)$ was defined in Eq. \eqref{WendrivenIsing}. The explicit form of the Hamiltonian in the rotating frame is given by
\begin{widetext}
\begin{eqnarray}
      \label{AppHamiltonianRotFrameSI}
	    \hat{H}_{i^{*},m}^{\sigma}(t)&=& -\Delta_{m}\sum_{j^{*}=1}^{N}\sigma_{i^{*},j^{*}}^x-\frac{J}{2}\left\{1+\cos[4\theta_{m}(t)]\right\}\sum_{j^{*}=1}^N \sigma_{i^{*},j^{*}}^z \sigma_{i^{*},j^{*}+1}^z	   
	    -\frac{J}{2}\left\{1-\cos[4\theta_{m}(t)]\right\}\sum_{j^{*}=1}^N \sigma_{i^{*},j^{*}}^y \sigma_{i^{*},j^{*}+1}^y
	    \nonumber \\&&
            +\frac{J}{2}\sin[4\theta_{m}(t)]\sum_{j^{*}=1}^N \sigma_{i^{*},j^{*}}^z \sigma_{i^{*},j^{*}+1}^y            
            +\frac{J}{2}\sin[4\theta_{m}(t)]\sum_{j^{*}=1}^N \sigma_{i^{*},j^{*}}^y \sigma_{i^{*},j^{*}+1}^z
     ,
\end{eqnarray}
\end{widetext}
where $\Delta_{m}=g_{0}-m (\Omega/4)$. By using the identity
\begin{equation}
       \label{ExpBessel} 
	     \exp(iz\sin\Omega t)=\sum_{l=-\infty}^{\infty}\mathcal{J}_{l}(z)\exp(il\Omega t) 
       ,
\end{equation}
where $\mathcal{J}_{l}(z)$ is the  $l$th-order Bessel function \cite{Abramowitz},
the Hamiltonian Eq. \eqref{AppHamiltonianRotFrameSI} can be written in the form
\begin{equation}
      \label{IsingHamiltonianFourier}
             \hat{H}^{\sigma}_{i^{*},m}(t)=\sum_{n=-\infty}^{\infty} h^{(\sigma,m)}_{i^{*},n}\exp{(i n \Omega t)}     
   .
\end{equation}
In analogy with the standard RWA of quantum optics, we obtain an approximate Hamiltonian to describe the $m$th resonance by neglecting all the terms in $\hat{H}_{m}(t)$ with oscillatory time-dependence:  $\hat{H}_{i^{*},m}^{\sigma}(t) \approx h^{(\sigma,m)}_{i^{*},0}$. This approximation is valid as long as the conditions Eq. \eqref{ValidityRWAIsing} hold. Finally, we obtain the time-independent effective Hamiltonian Eq. \eqref{mResonaceIsingXY}.

\section{Cycle-averaged correlation functions: Ac-driven long-range magnetic order \label{AppendixB}}
In this section we perform the calculation of the spin correlation functions in nonequilibrium of the $i^{*}$-th Ising chain Hamiltonian Eq. \eqref{WendrivenIsing}.

In particular, we are interested in the spin spatial correlations
\begin{equation}
      \label{SpinCorrelator}
            \eta_{\alpha \alpha}(R,t)=\left\langle\Phi^{(-)}_{m,i^{*}}(t)\left|\sigma_{1,1}^{\alpha}\sigma_{1,1+R}^{\alpha}\right|\Phi^{(-)}_{m,i^{*}}(t)\right\rangle          
\end{equation}
for $\alpha \in \{y,z\}$, where
\begin{equation}
      \label{AppIsingFloquetstate}
            |\Phi^{(-)}_{m,i^{*}}(t)\rangle 
            = \exp(-iN E^{(-)}_{m}t)\hat{U}_{i^{*},m}(t)|\chi^{(-)}_{m}\rangle       
\end{equation}
is the negative-quasienergy Floquet eigenstate. 

In the last expression, $E^{(-)}_{m}=-\int_0^{\pi}\frac{dk}{2\pi}\varepsilon_{k,m}$ is the total quasienergy, and $\varepsilon_{k,m}$ is the quasienergy 
dispersion defined in Eq. \eqref{QuasiEnergyExcitation}. Furthermore, the state $|\chi^{(-)}_{m}\rangle$ is defined in Eq. \eqref{Rot} and the  transformation $\hat{U}_{i^{*},m}(t)$ is given by Eq. \eqref{AppUnitaryIsingRot} .

Without loss of generality, let us perform here the explicit calculation for the particular case $\alpha=z$:
\begin{widetext}
\begin{eqnarray}
      \label{FinalCorrZZ}
	    \eta_{zz}(R,t)&=&(1+\cos[4\theta_{m}(t)])
	    \left\langle\chi^{(-)}_{m}\left|\sigma_{1,1}^{z}\sigma_{1,1+R}^{z}\right|\chi^{(-)}_{m}\right\rangle+	   
	    (1-\cos[4\theta_{m}(t)])
	    \left\langle\chi^{(-)}_{m}\left|\sigma_{1,1}^{y}\sigma_{1,1+R}^{y}\right|\chi^{(-)}_{m}\right\rangle
	    \nonumber \\&&
            +\sin[4\theta_{m}(t)]\left\langle\chi^{(-)}_{m}\left|\sigma_{1,1}^{z}\sigma_{1,1+R}^{y}\right|\chi^{(-)}_{m}\right\rangle            
            +\sin[4\theta_{m}(t)]\left\langle\chi^{(-)}_{m}\left|\sigma_{1,1}^{y}\sigma_{1,1+R}^{z}\right|\chi^{(-)}_{m}\right\rangle
     .
\end{eqnarray}
\end{widetext}
After averaging over a cycle we get
\begin{widetext}
\begin{equation}
      \label{CycleAvFinalCorrZZ}
           \frac{1}{T}\int_{0}^{T}\eta_{zz}(R,t)=\left[1+(-1)^{m}\mathcal{J}_{m}\left(\frac{4g_{1}}{\Omega}\right)\right]
           \left\langle\chi^{(-)}_{m}\left|\sigma_{1,1}^{z}\sigma_{1,1+R}^{z}\right|\chi^{(-)}_{m}\right\rangle+
           \left[1-(-1)^{m}\mathcal{J}_{m}\left(\frac{4g_{1}}{\Omega}\right)\right]
           \left\langle\chi^{(-)}_{m}\left|\sigma_{1,1}^{y}\sigma_{1,1+R}^{y}\right|\chi^{(-)}_{m}\right\rangle
      .
\end{equation}
\end{widetext}

\end{document}